\documentclass[10pt, twocolumn, apacite]{extarticle}

\usepackage{titling}
\usepackage{booktabs} 
\usepackage{siunitx} 
\usepackage{graphicx}
\usepackage{amsmath}
\usepackage{amssymb}
\usepackage{caption}

\usepackage{biblatex}
\newcommand{\func}[1]{\mathrm{#1}}

\newcommand{\N}{\mbox{\rm \hbox{I\kern-.15em\hbox{N}}}}

\newcommand{\of}[1]{\!\left( #1 \right)}
\newcommand{\setb}[1]{\!\left\{ #1 \right\}}

\newcommand{\norm}[1]{\left\Vert {#1} \right\Vert}

\usepackage[ruled]{algorithm2e} 

\SetAlFnt{\small}
\SetAlCapFnt{\small}
\SetAlCapNameFnt{\small}
\SetAlCapHSkip{0pt}

\begin{document}
\title{Neural Volumetric Blendshapes: Computationally Efficient Physics-Based Facial Blendshapes}

\author{Nicolas Wagner,Ulrich Schwanecke, and Mario Botsch}

\twocolumn[
\maketitle 
    \begin{abstract}
        Computationally weak systems and demanding graphical applications are still mostly dependent on linear blendshape models for facial animations. At this, artifacts such as self-intersections, loss of volume, or missing soft tissue elasticity are often avoided by using comprehensively designed blendshape rigs. However, hundreds of blendshapes have to be manually created or scanned in for high-quality animations, which is very costly to scale to many characters. Non-linear physics-based animation systems provide an alternative approach which avoid most artifacts by construction.  Nonetheless, they are cumbersome to implement and require immense computational effort at runtime. We propose neural volumetric blendshapes, a realtime approach on consumer-grade CPUs that combines the advantages of physics-based simulations while the handling is as effortless and fast as that of linear blendshapes. To this end, we present a neural network that efficiently approximates volumetric simulations and generalizes across human identities as well as facial expressions. Furthermore, it only requires a single neutral face mesh as input in the minimal setting. Along with the design of the network, we introduce a pipeline for the challenging creation of anatomically and physically plausible training data. Part of the pipeline is a novel layered head model that densely positions the biomechanical anatomy within a skin surface while avoiding intersections. The fidelity of all parts of the data generation pipeline as well as the accuracy and efficiency of the network are evaluated in this work. Upon publication, the trained models and associated code will be released.
    \end{abstract}
    \vspace{0.5cm}
]

\begin{figure*}
\centering
\includegraphics[width=\textwidth]{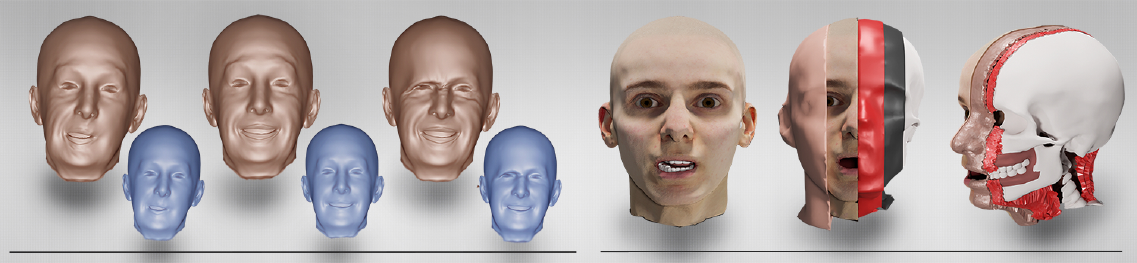}
\vspace{-0.8cm}\caption*{\hspace{-0.0cm} a) \hspace{9.2cm} b)}\vspace{-0.3cm}
\caption{a) Exemplary results of our Neural Volumetric Blendshapes (brown) compared to linear blendshapes (blue). Among others, volume preservation, the ability to create more detailed wrinkles, and avoided self-intersections result in more realistic and anatomically plausible facial animations. b) An example fit of the novel layered head model. The model encapsulates the skin, the skull, and the muscles with wraps for which we present a data-driven fitting algorithm. The space between the wraps can be canonically tetrahedralized.}
\end{figure*}

\section{Introduction}
At present, research in the field of head avatars and facial animation is mainly concerned with obtaining photorealistic results through neural networks \cite{cao2022authentic, grassal2022neural, athar2022rignerf} which can be operated on computationally rich systems, require comprehensive per-person training data, and time-consuming individualization. What currently falls short, however, is the inclusion of less capable hardware setups and efficient training pipelines that avoid extensive data collection. For this, various adaptions of linear blendshape models paired with deformation transfer \cite{sumner2004deformation, botsch2006deformation} and example-based facial rigging \cite{li2010example} are still the usual means in production. Although linear facial models have been intensively researched and improved over the past decades, there are still known shortcomings like physically implausible distortions, loss of volume, anatomically impossible expressions, missing volumetric elasticity, or self-intersections. Physics-based simulations have been proposed that overcome most artifacts of linear blendshapes \cite{ichim2017phace, ichim2016building, cong2016art, choi2022animatomy}, but they are usually laborious to handle and computationally expensive. Hybrid models that try to combine the best of both worlds are either not sophisticated enough in the quality of the simulated physical properties \cite{barrielle2016blendforces} or still too inefficient to be used on slower devices \cite{ichim2016building}.

An auspicious approach in the latter category is physics-based volumetric blendshapes \cite{ichim2016building}. These can be animated with physical plausibility, anatomical constraints can be taken into account, self-intersections can be prevented, and the control is identical to linear blendshapes. Although the level of detail is slightly sacrificed in comparison to other physics simulations \cite{cong2016art, ichim2017phace}, volumetric blendshapes can still only be used at low frame rates. We improve on this approach with \emph{neural volumetric blendshapes} that approximate the involved calculations of physical and anatomical constraints with an efficient and lightweight neural network. Thereby, realtime inference of physics-based non-linear blendshapes on consumer-grade hardware becomes possible and only a slight computational overhead is necessary compared to linear blendshapes. The principal challenge we solve in this work is the creation of training data for neural volumetric blendshapes that reflects the previously discussed animation advantages and facilitates a generalization across different human identities. Thus, in contrast to other recent work that tries to approximate physics-based facial simulations by neural networks \cite{srinivasan2021learning, yang2022implicit, choi2022animatomy}, we avoid time-consuming individualizations for a straightforward deployment. In addition, our facial animation method does not require sequences of optical scans or manually crafted facial animations. Instead, the network trained in this work induces an anatomical plausible deformation transfer such that our system is instantly applicable to neutral head surfaces or on top of arbitrary linear blendshape rigs.

The aforementioned data generation pipeline is unique in that, to the best of our knowledge, there are no other comprehensive datasets that relate a broad range of head shapes in diverse facial expressions to the underlying anatomical and physical characteristics. To curate such a dataset for the first time, we bring together multiple data sources such as CT data to reflect the anatomy of heads, 3D reconstructions of images in the wild to collect diverse head shapes, or facial expressions in the form of recorded blendshape weights from dyadic conversational situations. The result is a dataset of heads with neutral and non-neutral expressions represented by a novel standardized layered head model, relating skin surface displacements with the underlying biomechanical volumetric deformations and transformations of muscles, skull bones, and soft tissue.

The \emph{key novelties and contributions} we present in this paper can be summarized as follows:
\begin{itemize}
    \item A novel \emph{layered head model} (LHM) representing the skin surface as well as the entire biomechanical anatomy.
    \item A data-driven procedure for \emph{fitting the LHM} to neutral skin surfaces.
    \item An \emph{inverse physics-based simulation} which fits the LHM to skin surfaces of non-neutral facial expressions.
    \item A novel neural network design that approximates physics-based simulations to efficiently implement neural volumetric blendshapes.
    \item A pipeline for \emph{creating training data} of the neural network that among other parts includes the LHM and the associated fitting procedures.
\end{itemize}

\section{Related Work}
\subsection{Personalized Anatomical Models}
Algorithms that create {personalized anatomical models} can essentially be distinguished according to two paradigms: \emph{heuristic-based} and \emph{data-driven}. Considering heuristic-based approaches, Anatomy Transfer \cite{ali2013anatomy} applies a space warp to a template anatomical structure to fit a target skin surface. The skull and other bones are only deformed by an affine transformation. A similar idea is proposed by Gilles et al. \cite{gilles2010creating}. While they also implement a statistical validation of bone shapes, the statistics are collected from artificially deformed bones. In \cite{ichim2016building, kadlevcek2016reconstructing}, an inverse physics simulation was used to reconstruct anatomical structures from multiple 3D expression scans. Saito et al.~\cite{saito2015computational} simulate the growth of soft tissue, muscles, and bones. A musculoskeletal biomechanical model is fitted from sparse measurements in \cite{schleicher2021bash} but not qualitatively evaluated.

There are only a few data-driven approaches because combined data sets of surface scans and MRI, CT or DXA images are hard to obtain for various reasons (e.g. data privacy or unnecessary radiation exposure). The recent work OSSO \cite{keller2022osso} predicts full body skeletons from 2000 DXA images that do not carry precise 3D information. Further, bones are positioned within a body by predicting only three anchor points per bone group and not avoiding intersections between skin and skull. A model that prevents skin-skull intersections and also considers muscles is based on fitting encapsulating wraps instead of the anatomy itself \cite{komaritzan2021inside}. However, no accurate algorithm based on medical imaging but a BMI (Body mass index) regressor \cite{maalin2021beyond} is used to position the wraps. A much more accurate, pure face model, was developed by Achenbach et al.~\cite{achenbach2018multilinear}. Here, CT scans are combined with optical scans by a multilinear model (MLM) which can map from skulls to faces and vice versa. As before, no self-intersections are prevented and only bones are fitted. Building on the data from \cite{achenbach2018multilinear} and following the idea of a layered body model \cite{komaritzan2021inside}, we create a statistical layered head model including musculature that avoids self-intersections.

\subsection{Physics-Based Facial Animation}
A variety of techniques for animating faces have been developed in the past \cite{ichim2015dynamic, bradley2010high, zhang2008spacetime, parke1991control}. Data-driven models \cite{lewis2014practice, ichim2016building, lewis2005reducing}, which have recently been significantly improved by deep learning \cite{zheng2022avatar, garbin2022voltemorph, cao2022authentic, feng2021learning, song2020accurate, athar2022rignerf}, are certainly dominant. Due to their simplicity and speed, linear blendshapes \cite{lewis2014practice} are still most commonly used in demanding applications and whenever no computationally rich hardware is available. Physics-based models have been developed for a long time \cite{sifakis2005automatic} and avoid artifacts like implausible contortions and self-intersections, but due to their complexity and computational effort, they are rarely used. Hybrid approaches add surface-based physics to linear blendshapes for more detailed facial expressions \cite{barrielle2016blendforces, bickel2008pose, choi2022animatomy}. However, by construction they can not model volumetric effects.

The pioneering work of Sifakis et al.~\cite{sifakis2005automatic} is the first fully phy\-sics-based facial animation. The simulation is conducted on a personalized tetrahedron mesh, which can only be of a limited resolution due to a necessary dense optimization problem. With Phace \cite{ichim2017phace}, this problem was overcome by an improved physics simulation. An art-directed muscle model \cite{cong2019muscle, bao2019high, cong2016art} additionally represents muscles as B-splines and allows control of expressions via trajectories of spline control points. A solely inverse model for determining the physical properties of faces was proposed in \cite{kadlevcek2019building}. Neural soft-tissue dynamics \cite{santesteban2020softsmpl, casas2018learning} extend the SMPL (Skinned Multi-Person Linear Model) proposed in \cite{SMPL:2015} with secondary motion. Recently, \cite{srinivasan2021learning, yang2022implicit, choi2022animatomy} adapted neural soft-body dynamics to learn the physical properties of a particular person. However, these approaches must be retrained for new objects and are slow in inference. With volumetric blendshapes \cite{ichim2016building}, a hybrid approach has been presented that combines the structure of linear blendshapes with physical and anatomical plausibility. We extend this work to \emph{neural volumetric blendshapes} to make physical plausibility realtime capable while maintaining the control structure of standard linear blendshapes.

\section{Method}
The foundation of our neural volumetric blendshapes approach (Sections \ref{sec:vbs} and \ref{sec:nnvbs}) is a novel layered head representation (Section \ref{sec:lhm}). Starting from there, we design a physics-based facial animation model (Section \ref{sec:pbm}) and distill it into a defining dataset (Section \ref{sec:pip}). With this dataset, a neural network that makes the animation model realtime capable is trained.

\subsection{Layered Head Model}
\begin{figure}[t]
\includegraphics[width=1.0\linewidth]{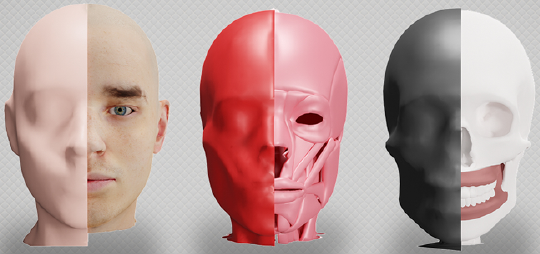}\centering
\caption{All components of the layered head model template $\mathcal{T}$. Skin $S_{\mathcal{T}}$, skin wrap $\hat{S}_{\mathcal{T}}$, muscles $M_{\mathcal{T}}$, muscles wrap $\hat{M}_{\mathcal{T}}$, skull $B_{\mathcal{T}}$, and the skull wrap $\hat{B}_{\mathcal{T}}$. Each wrap is an abstract and simplified representation of a more complex structure.}
\label{fig:temp}
\end{figure}
\label{sec:lhm}
We represent a head $\mathcal{H} = \rho_{\mathcal{H}}\of{\mathcal{T}}$ with neutral expression through a component-wise transformation (see Section \ref{sec:pip} for details) of a layered head model (LHM) template
\begin{equation}
\mathcal{T} = \of{S_{\mathcal{T}}, B_{\mathcal{T}}, M_{\mathcal{T}}, \hat{S}_{\mathcal{T}}, \hat{B}_{\mathcal{T}}, \hat{M}_{\mathcal{T}}},
\end{equation}
that consists of six triangle meshes. $S_{\mathcal{T}}$ describes the skin surface including the eyes, the mouth cavity, and the tongue, $B_{\mathcal{T}}$ the surface of all skull bones and teeth, $M_{\mathcal{T}}$ the surface of all muscles and the cartilages of the ears and nose. $\hat{S}_{\mathcal{T}}$ is the skin layer, i.e.\ a closed wrap enveloping $S_{\mathcal{T}}$, $\hat{B}_{\mathcal{T}}$ the skull layer that envelopes $B_{\mathcal{T}}$, and $\hat{M}_{\mathcal{T}}$ the muscle layer that envelopes $M_{\mathcal{T}}$. Other anatomical structures are omitted for simplicity. 
The template structures $S_{\mathcal{T}}, B_{\mathcal{T}}$ and $M_{\mathcal{T}}$ were designed by an experienced digital artist. The skin, skull, and muscle layers $\hat{S}_{\mathcal{T}}, \hat{B}_{\mathcal{T}}$, and $\hat{M}_{\mathcal{T}}$ have the same triangulation and were generated by shrink-wrapping a sphere as close as possible to the corresponding surfaces without intersections. The complete template is shown in Figure \ref{fig:temp}.

The skull and muscle layers are further massaged such that the quads of all prisms that can be spanned between corresponding faces of the skin, muscle, and skull layer are as rectangular as possible while preserving the original geometries. To that end, we determine for the skull layer
\begin{equation}\label{eq:initial}
\begin{aligned}
     \arg \min_{X} \hspace{1pt} 
     & w_\mathrm{rect} {E}_\func{rect}\of{X, \hat{S}_{\mathcal{T}}} +
     w_\mathrm{dist} E_\func{dist}\of{X, \hat{B}_{\mathcal{T}}}, 
\end{aligned}
\end{equation}
within the efficient projective dynamics \cite{bouaziz2014projective} optimization framework, initializing with $X = \hat{B}_{\mathcal{T}}$. Here, $E_{d}$ is the two-sided Hausdorff distance to the non-massaged shape and 
\begin{equation}
\begin{aligned}\label{eq::rect}
    & E_{r}\of{X, \hat{S}_{\mathcal{T}}} = \\
    & \sum_{(x_i^0, x_i^1) \in X} \left(90^{\circ}  - \angle \of{x_i^0 x_i^1 \hat{s}_i^1}\right)^2 
    +\left(90^{\circ} - \angle \of{x_i^1 x_i^0 \hat{s}_i^0}\right)^2 \\
    & \quad\quad\quad +\left(90^{\circ} - \angle \of{\hat{s}_i^0 \hat{s}_i^1 x_i^1}\right)^2
    +\left(90^{\circ} - \angle \of{\hat{s}_i^1 \hat{s}_i^0 x_i^0}\right)^2,
\end{aligned}
\end{equation}
induces the rectangular prism shapes.
After the optimization we set $\hat{B}_{\mathcal{T}} = X$. The same optimization is run for the muscle layer.

\begin{table}[]
\centering
\begin{tabular}{lllllllll}
\toprule
Mesh & $S_{\mathcal{T}}$ & $B_{\mathcal{T}}$  & $M_{\mathcal{T}}$  & $\hat{S}_{\mathcal{T}}$ \\
 \#Vertices & 21875 & 14572 &  16388 &  7826  \\
 \#Faces / \#Tets & 42738 &  28856&  32370 &  15648 \\
\bottomrule
&&&& \\
\toprule
Mesh &  $\hat{B}_{\mathcal{T}}$  & $\hat{M}_{\mathcal{T}}$  & $\mathbb{S}_{\mathcal{T}}$  & $\mathbb{M}_{\mathcal{T}}$\\
 \#Vertices  &  7826 &  7826 &  \multicolumn{2}{c}{49852}   \\
 \#Faces / \#Tets & 15648 &  15648&  123429 & 73681  \\
\bottomrule
\end{tabular}
\vspace{2.4mm}
\caption{Description of the cardinality of each template LHM $\mathcal{T}$ component. By subdividing the wrap meshes or the layer prisms, the resolution of the template tetrahedron meshes can easily be adjusted.}
\label{tab::sizes}
\end{table}

The wrapping layers of the LHM allow for at least two significant advantages. On the one hand, they provide a simplified representation of the skin surface, the musculature, and the skull, which we exploit in Section \ref{sec:pip} for determining the LHM template transformation $\rho_{\mathcal{H}}$.  On the other hand, they can be used for topologically and semantically consistent tetrahedralization of the head volume by splitting the prisms between the layers canonically into tetrahedrons.

More precisely, the construction of the LHM also defines a soft tissue tetrahedron mesh $\mathbb{S}_{\mathcal{T}}$ (i.e. between the skin and the muscle layer) and a muscle tissue tetrahedron mesh $\mathbb{M}_{\mathcal{T}}$ (i.e. between the muscle and the skull layer). The massage of the muscle and skull layers ensures nicely shaped tets with minimal shearing. Further, $\mathbb{S}_{\mathcal{T}}$ can be fine-tuned by removing all vertices and connected tets outside of $S_{\mathcal{T}}$, adding the vertices of $S_{\mathcal{T}}$, and connecting them via Delaunay tetrahedralization. The complexities of all template components are given in Table \ref{tab::sizes}.

\subsection{Linear \& Volumetric Blendshapes}\label{sec:vbs}
Building on the LHM representation, we can now introduce neural volumetric blendshapes. For this, the classical concept of linear blendshapes is reviewed first. Thereupon, we derive volumetric blendshapes and the involved physics-based simulations that are in general not real time capable. Finally, we introduce neural volumetric blendshapes (Section \ref{sec:nnvbs}) as an efficient and fast alternative.

For a specific head $\mathcal{H} = \of{S_{\mathcal{H}}, B_{\mathcal{H}}, M_{\mathcal{H}}, \hat{S}_{\mathcal{H}}, \hat{B}_{\mathcal{H}}, \hat{M}_{\mathcal{H}}}$, a linear blendshape model consists of $n$ surface blendshapes 
\begin{equation}\label{eq::lin_rig}
\setb{{S}_{\mathcal{H}}^{i}}_{i=1}^n
\end{equation} and determines an unknown facial expression $S_{\mathcal{H}}^\func{exp}$ as the linear interpolation 
\begin{equation}    
S_{\mathcal{H}}^\func{exp} = \sum_{i} w_\mathrm{exp}^i S_{\mathcal{H}}^i.
\end{equation}
 The $w_\mathrm{exp}^i$ are the blending weights and determine the share of each blendshape in the expression. 
 
The corresponding set of volumetric blendshapes can be defined as 
\begin{equation} 
    \left\{V^i_{\mathcal{H}}\right\}_{i=1}^n = \left\{\of{\nabla\mathbb{S}_{\mathcal{H}}^{i}, \nabla\mathbb{M}_{\mathcal{H}}^{i}, B_{\mathcal{H}}^i}\right\}_{i=1}^n,
\end{equation}
where $\nabla\mathbb{S}_{\mathcal{H}}^{i}$, $\nabla\mathbb{M}_{\mathcal{H}}^{i}$ describe soft and muscle tissue deformations as vectors of $3\times 3$ deformation gradients. $B_{\mathcal{H}}^i$ describes the skull after rigid motion of jaw and cranium. An anatomically plausible inverse physics model $\phi^{\dagger}$ (see Section \ref{sec:pbm} for details) relates the linear rig to the volumetric rig as
\begin{equation}\label{eq::link}
V^i_{\mathcal{H}} = \phi^{\dagger}\left(S_{\mathcal{H}}^{i}, \mathbb{S}_{\mathcal{H}}, \mathbb{M}_{\mathcal{H}}, B_{\mathcal{H}}\right).
\end{equation} 

In words, $\phi^{\dagger}$ determines the biomechanical volumetric deformations of $\mathcal{H}$ that cause the skin surface of a facial expression. Vice versa, a forward physics model $\phi$ acts as a left inverse to $\phi^{\dagger}$ and maps volumetric deformations back to surface deformations as
\begin{equation}
\bar{S}_{\mathcal{H}}^{i} = \phi\of{V^i_{\mathcal{H}}}.
\end{equation}

In reality, ${S}_{\mathcal{H}}^{i}$ is often not an anatomically legal facial expression with respect to $\phi^{\dagger}$. As a consequence, $\phi^{\dagger}$ is built such that $\bar{S}_{\mathcal{H}}^{i}$ is an anatomically reachable expression which, in Euclidean sense, is close but not necessarily equal to ${S}_{\mathcal{H}}^{i}$.
 
The major advantage of a volumetric blendshape model is that unknown facial expressions calculated as \begin{equation}\label{eq::eval_vbs}
    \bar{S}_{\mathcal{H}}^\func{exp} = \phi\of{\bigoplus_{i}w_\mathrm{exp}^i \otimes V^i_{\mathcal{H}}}
\end{equation}
can be shaped anatomically more plausible and thus more realistic, provided a suitable choice of interpolation (and extrapolation) functions $\bigoplus$ and $\otimes$. At this, the essential requirement to $\bigoplus$ and $\otimes$ is the biologically necessary volume preservation of the deformation gradients. The common approach would be to to separately interpolate the stretch and the rotation components of the deformation gradients \cite{ichim2016building, shoemake1992matrix} and the positions of the skull bones. Considering the stretching components, volume-preserving interpolation methods \cite{arsigny2007geometric, jung2015scaling} have been proposed. Considering the rotation components, quaternion-interpolation satisfies the volume-preservation by construction. However, to the best of our knowledge, it is yet to be discussed how both components should be extrapolated for facial animations if the blending weights do not form a convex combination. In this work, we therefore use a novel hybrid approach that calculates facial expressions as
\begin{equation}\label{eq::eval_nvbs_t}    
 \bar{S}_{\mathcal{H}}^\func{exp} = \phi\left(\phi^{\dagger}\of{\sum_{i} w_\mathrm{exp}^i \bar{S}_{\mathcal{H}}^i, \mathbb{S}_{\mathcal{H}}, \mathbb{M}_{\mathcal{H}}, B_{\mathcal{H}}}\right).
 \end{equation} 
This way, inter- and extrapolation capabilities of linear surface blendshapes can be used while maintaining the advantages of anatomical plausibility through $\phi^{\dagger}$.

\subsection{Neural Volumetric Blendshapes}\label{sec:nnvbs}
\begin{figure*}[t]
  \includegraphics[width=\linewidth]{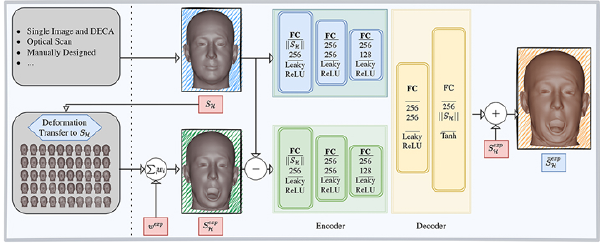}
  \caption{ An overview of the neural volumetric blendshape model and how novel expressions are determined. For the fully connected layers (FC), the input size, the output size, as well as the activation function are stated. Please note that $f$ is designed as an adapter such that the part left of the dotted line can be amended.}\label{fig:net}
\end{figure*}
Regardless of the construction of $\bigoplus$ and $\otimes$, the calculation of  facial expressions is in general not realtime-capable for high-resolution volumetric blendshapes and sophisticated physics models $\phi$ and $\phi^{\dagger}$. We therefore present the neural volumetric blendshape model that like the linear blendshape model \eqref{eq::lin_rig} consists of a set of (not necessarily anatomically plausible) expression surfaces $\{{S}_{\mathcal{H}}^{i}\}_{i=1}^n$. 
However, unknown \emph{anatomically plausible} expressions are formed by a neural network $f$ that is trained such that
 \begin{equation}\label{eq::eval_nvbs}    
 \bar{S}_{\mathcal{H}}^\func{exp} - \sum_{i} w_\mathrm{exp}^i S_{\mathcal{H}}^i \approx f\of{\sum_{i} w_\mathrm{exp}^i S_{\mathcal{H}}^i - S_{\mathcal{H}}, S_{\mathcal{H}}}.
 \end{equation}
The structure of Equation \eqref{eq::eval_nvbs} enables us to implement $f$ efficiently as fully connected networks. More precisely, both inputs to $f$, the neutral surface $S_{\mathcal{H}}$ and the vector of the differences to the linear interpolation result $S_{\mathcal{H}}^\func{exp} = \sum_{i} w_\mathrm{exp}^i S_{\mathcal{H}}^i$, are tokenized with the help of respective encoders. Subsequently, the tokens are processed by a decoder that outputs the vertex-wise deformations from $S_{\mathcal{H}}^\func{exp}$ to the anatomically plausible expression $\bar{S}_{\mathcal{H}}^\func{exp}$.

The inputs and outputs of $f$ are justified as discussed next. Considering Equation \eqref{eq::eval_nvbs_t} which is approximated by $f$, it seems reasonable to use the same inputs and only learn the evaluation of $\phi(\phi^{\dagger}(\cdot)$. However, the accompanying tetrahedron meshes are significantly higher-dimensional than the corresponding surface meshes and $f$ would be slow downed severely in the inference speed. We therefore expect $f$ to implicitly learn the linking between linear and volumetric blendshapes as shown in Equation \eqref{eq::link} as well as the fitting of the LHM. Since we demonstrate in Section \ref{sec:pip} how the LHM can be fitted given only the neutral surface and also the linking does not need any further inputs, the neutral surface carries sufficient information to omit the tetrahedron meshes. For the second input, the difference vector, the blending weights or the linear interpolation result could alternatively be inserted. Inputting only the weights, however, would considerably limit the flexibility of $f$ because it does not allow the underlying surface blendshapes to be changed after training. On the other hand, inputting the linear interpolation result exhibits disadvantages at training time. The target deltas are mostly Gaussian distributed and can therefore be learned more easily by a neural network \cite{li2020dynamic}. In the same spirit, the output of $f$ is chosen to be the differences to the anatomically plausible expression.

We evaluated alternatives to fully connected networks such as set transformers \cite{lee2019set}, convolutional networks on geometry images, graph neural networks \cite{scarselli2008graph}, or implicit architectures \cite{mildenhall2021nerf}, but all have exhibited substantially slower inference speeds while reaching a similar accuracy.

Our design of $f$ is not only fast, i.e.,\ runs only 8 ms per frame on a consumer grade Intel i5 12600K, but it is also straightforward to deploy. A single neutral surface of a head is sufficient on which deformation transfer \cite{botsch2006deformation} can be applied. Building on this, an anatomically plausible animation can be performed with $f$ as shown in Figure \ref{fig:net}. Further, no more volumetric information has to be processed in complex simulation frameworks. Thereby, and because only simple fully connected layers are used, the mesh is portable and easy to use on many different (computationally weak) devices.

Next, the construction of $\phi$ and $ \phi^{\dagger}$ is described and subsequently a pipeline for creating a dataset to learn $f$ is structured. 

\begin{figure*}[t]
  \centering
  \includegraphics[width=1.\textwidth]{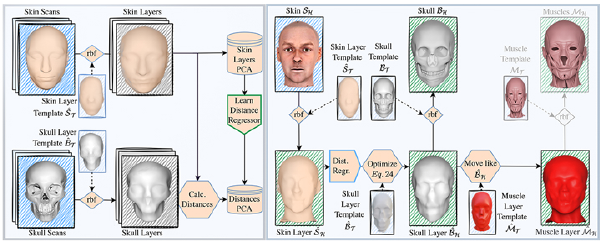}
  \vspace{-0.8cm}\caption*{\hspace{-0.5cm} a) \hspace{8.2cm} b)}\vspace{-0.3cm}
  \caption{a) The training scheme of the skin to skull layer distances regressor $D$ which is used in the layered head model fitting. $D$ is trained on the SKULLS dataset \cite{gietzen2019method} that relates optical skin scans to MRI measurements. b) Procedural overview of the layered head model fitting algorithm. Starting from only a neutral head scan, the five other components are positioned. Blue strokes indicate inputs, green strokes outputs.}
\label{fig:fitting}
\end{figure*}

\subsection{Physics-Based Simulations}\label{sec:pbm}
We realize the anatomically plausible inverse physics $\phi^{\dagger}$ and the left inverse $\phi$ as projective dynamics energies $E_{\phi^{\dagger}}$ and $E_{\phi}$, respectively. Starting with $\phi^{\dagger}$, separate terms for soft tissue, muscle tissue, the skin, the skull, and auxiliary components are applied. Considering the soft tissue $\mathbb{S}_{\mathcal{H}}$, we closely follow the model of \cite{ichim2017phace} and impose
\begin{equation}
\label{eq::energy_soft}
    E_{{\mathbb{S}_{\mathcal{H}}}}  = 
    w_\mathrm{vol} \sum_{t \in \mathbb{S}_{\mathcal{H}}} E_\mathrm{vol}(t)  +
    w_\mathrm{str} \sum_{t \in \mathbb{S}_{\mathcal{H}}} E_\mathrm{str}(t),
\end{equation}
which for each tetrahedron $t$ penalizes changes of volume
\begin{equation}
    \begin{aligned}
    E_\mathrm{vol}(t) = \of{\det(\text{F}(t)) - 1}^2
    \end{aligned}
\end{equation}
and strain
\begin{equation}\label{eq:strain}
    \begin{aligned}
    E_\mathrm{str}(t) = \min_{R \in SO(3)} \left\Vert \text{F}(t) - R\right\Vert^2_{F}.
    \end{aligned}
\end{equation}
$\text{F}(t)$ denotes the deformation gradient of a tetrahedron $t$, $R \in SO(3)$ the optimal rotation, and $\left\Vert\cdot\right\Vert^2_{F}$ the Frobenius norm. 

To reflect the biological structure of the skin, we additionally formulate a dedicated strain energy
\begin{equation}
    E_{{S_{\mathcal{H}}}}  = 
    \sum_{t \in {S}_{\mathcal{H}}}  E_\mathrm{str}(t)
\end{equation} on each triangle $t$ of the skin.

For the muscle tets $\mathbb{M}_{\mathcal{H}}$, we follow the arguments of \cite{kadlevcek2019building} that capturing fiber directions for tetrahedralized muscles is in general too restrictive. Hence, only a volume-pre\-ser\-vation term
\begin{equation}
    E_{{\mathbb{M}_{\mathcal{H}}}}  = w_\mathrm{vol}\sum_{t \in \mathbb{M}_{\mathcal{H}}}  E_\mathrm{vol}(t)
\end{equation} 
is applied. The weight $w_\mathrm{vol}$ is set sufficiently high such that the volume-preservation is almost a hard constraint.

The skull is not tetrahedralized as it is assumed to be non-de\-form\-able even though it is rigidly movable. The non-deformability of the skull is represented by
\begin{equation}
    E_{B_{\mathcal{H}}} \!= 
     \sum_{t \in B_{\mathcal{H}}}\! E_\mathrm{str}\of{t} + 
     \sum_{x \in B_{\mathcal{H}}}\! E_\mathrm{curv}\of{x, B_{\mathcal{H}}},
\end{equation}
i.e.\ a strain $E_{str}$ on the triangles and mean curvature regularization
\begin{equation}\label{eq::bend}
    E_\mathrm{curv}\of{x, B_{\mathcal{H}}} =  A_x\norm{\Delta x - R \Delta b_x}^2
\end{equation}
on the vertices of the skull.
The matrix $R \in SO(3)$ denotes the optimal rotation keeping the vertex Laplacian $\Delta x$ as close as possible to its initial value $\Delta b_x$. The vertex Laplacian is discretized using the cotangent weights and  the Voronoi areas $A_x$ \cite{botsch2010polygon}. We do not model the non-deformability as a rigidity constrain due to the significantly higher computational burden.
Like for the volume constraints, the non-deformability of the skull is weighted to be an almost hard constraint.

To connect the muscle tets as well as the eyes to the skull, connecting tets are introduced. For the muscle tets, each skull vertex connects to the closest three vertices in $\mathbb{M}_{\mathcal{H}}$ to form a connecting tet. For the eyes, connecting tets are formed by connecting each eye vertex to the three closest vertices in $B_{\mathcal{H}}$. On these connecting tets, the energy $E_{con}$ with the same constraints as in Equation \eqref{eq::energy_soft} is imposed as almost hard constraints. By this design, the jaw and the cranium are moved independent from each other though muscle activations but the eyes remain rigid and move only with the cranium. 

Finally, the energy 
\begin{equation}
    E_\mathrm{inv} = \sum_{x \in S_{\mathcal{H}}} E_\mathrm{tar}\of{x, {S}^\func{exp}_{\mathcal{H}}}
\end{equation} 
of soft Dirichlet constraints 
\begin{equation}
    \begin{aligned}
        E_\mathrm{tar}\of{x, {S}^\func{exp}_{\mathcal{H}}} = \norm{ x - {s_x}}^2,
    \end{aligned}
\end{equation} 
is added, attracting each vertex $x$ of the skin surface $S_{\mathcal{H}}$ to the corresponding vertex $s_x$ from the target expression ${S}^\func{exp}_{\mathcal{H}}$.
As previously mentioned, the expression ${S}^\func{exp}_{\mathcal{H}}$ might be anatomically implausible. As a countermeasure, we can impose a maximum strain by balancing $w_\mathrm{str}$ with $w_\mathrm{tar}$. Thus, together with the almost hard constraints and by the construction of projective dynamics, $\phi^{\dagger}$ always results in a plausible expression $\bar{S}^\func{exp}_{\mathcal{H}}$ close to the target.
The weighted sum of the aforementioned energies gives the total energy 
\begin{equation}
\begin{aligned}
    E_{\phi^{\dagger}} = 
    & \:\: w_\mathrm{{\mathbb{S}_{\mathcal{H}}}}E_{{\mathbb{S}_{\mathcal{H}}}} 
      + w_\mathrm{{\mathbb{M}_{\mathcal{H}}}}E_{{\mathbb{M}_{\mathcal{H}}}} 
      + w_\mathrm{{B_{\mathcal{H}}}}E_{{B_{\mathcal{H}}}}\\
    & + w_\mathrm{{{S}_{\mathcal{H}}}}E_{{{S}_{\mathcal{H}}}} + w_\mathrm{con}E_{con} + w_\mathrm{inv}E_{inv}
\end{aligned}
\end{equation}
of the backward model $\phi^{\dagger}$.

The forward model $\phi$ is considerably simpler in structure and is realized as the energy
\begin{equation}\begin{aligned}
    E_{\phi}= &{}  \sum_{t \in \mathbb{S}_{\mathcal{H}}} E_\mathrm{dg}\of{t, \nabla\mathbb{S}_{\mathcal{H}}^\func{exp}} + \sum_{t \in \mathbb{M}_{\mathcal{H}}} E_\mathrm{dg}\of{t, \nabla\mathbb{M}_{\mathcal{H}}^\func{exp}}  \\
    &+ w_\mathrm{tar}\sum_{x \in B_{\mathcal{H}}} E_\mathrm{tar}\of{x, {B}^\func{exp}_{\mathcal{H}}} + w_\mathrm{con}E_\mathrm{con} \\
\end{aligned}
\end{equation} 
that aims to achieve given deformation gradients $\of{\nabla\mathbb{S}_{\mathcal{H}}^\func{exp}, \nabla\mathbb{M}_{\mathcal{H}}^\func{exp}}$ while matching the positions of the targeted skull ${B}^\func{exp}_{\mathcal{H}}$. Here, 
\begin{equation}
    \begin{aligned}
    E_\mathrm{dg}(t, \nabla\mathbb{T}) = \min_{R \in SO(3)} \norm{ \text{F}(t) - RT_t}^2_{F}
    \end{aligned}
\end{equation}
is similar to Equation \eqref{eq:strain} and attracts the deformation of each tetrahedron $t$ to a corresponding target deformation gradient $T_t \in \nabla\mathbb{T}$.

For both models we resolve self-intersections between colliding lips or teeth in a subsequent projective dynamics update. In the second update, colliding vertices are resolved as in \cite{komaritzan2018projective}. The distinctive feature here is that no collision-gaps can occur after dissolving the self-intersections.

\subsection{Generation of Training Data}\label{sec:pip}
To approximate Equation \eqref{eq::eval_nvbs_t} with $f$, a defining training dataset is required in first place. By the construction of $f$, this training dataset must consist of instances that relate diverse facial expressions created through linear blendshapes to the corresponding anatomically plausible surfaces. This dataset must also cover a variety of distinct head shapes to train $f$ to be as generally applicable as $\phi$ and $\phi^{\dagger}$.

In the following, we describe a pipeline for creating instances of such dataset, which can be roughly divided into two high-level steps. First, in order to evaluate Equation \eqref{eq::eval_nvbs_t} on a reasonable domain, it is necessary to model a head $\mathcal{H}$ from an extensive head-model and determine the corresponding neutral soft and muscle tissue tetrahedron meshes $\mathbb{S}_{\mathcal{H}}$ and $\mathbb{M}_{\mathcal{H}}$. Second, $S_{\mathcal{H}}$ has to be deformed to an expression ${S}_{\mathcal{H}}^\func{exp}$ and mapped to the anatomically plausible $\bar{S}_{\mathcal{H}}^\func{exp}$. A more detailed algorithmic description is given in Algorithm 1.

\textbf{Sampling Head Shapes} We start the first part of the pipeline by randomly drawing a neutral skin surface $S_{\mathcal{H}}$ from DECA \cite{feng2021learning}, one of the most comprehensive high-resolution face models currently available. More specifically, we randomly draw an image from the Flickr-Faces-HQ \cite{karras2019style} dataset and let DECA determine the corresponding neutral head shape. Further, a precomputed mapping is applied to adapt the DECA topology to our template.

\begin{table}[t]
\begin{tabular}{p{0.1\linewidth}p{0.8\linewidth}}
\toprule
\multicolumn{2}{l}{ Algorithm 1 \textbf{Data Generation}} \\ \midrule
\multicolumn{2}{l}{\textbf{Head Sampling and LHM Fitting }}  \\ 
\textit{1. a)} & Draw a random image $I_{\mathcal{H}}$ of a head $\mathcal{H}$ from the \newline FlickerHQ dataset. \\
\textit{\hphantom{1.} b)} & Calculate skin surface $S_{\mathcal{H}} = \mathit{DECA}\of{I_{\mathcal{H}}}$ with neutral expression parameters. \\
\textit{\hphantom{1.} c)} & Find LHM transformation $\rho_{\mathcal{H}}$ from $S_{\mathcal{H}}$, build up tetrahedron meshes $\mathbb{S}_{\mathcal{H}}$ and $\mathbb{M}_{\mathcal{H}}$. \\

\multicolumn{2}{l}{ \rule{0pt}{2.5ex}\textbf{Expression Sampling and Simulation }}  \\ 
\textit{2. a)} & Create ARKit blendshapes $\{S_{\mathcal{H}}^{i}\}_{i=1}^{52}$ from $S_{\mathcal{H}}$ with deformation transfer. \\
\textit{\hphantom{2.} b)} & Sample weights $w_\mathrm{exp} = \{w^i_\mathrm{exp}\}_{i=1}^{52}$ from dyadic recordings and calculate $S_{\mathcal{H}}^\func{exp} = \sum_{i=1}^{52} w^i_\mathrm{exp} S_{\mathcal{H}}^{i}$. \\
\rule{0pt}{2.5ex}%
\textit{\hphantom{1.} c)} & Get $\bar{S}_{\mathcal{H}}^\func{exp} $ from $ \phi\of{\phi^{\dagger}\of{S_{\mathcal{H}}^\func{exp}, \mathbb{S}_{\mathcal{H}}, \mathbb{M}_{\mathcal{H}}, B_{\mathcal{H}}}}$.\\ 
\bottomrule
\end{tabular}
\end{table} 

\textbf{Fitting the LHM} Next, the template LHM $\mathcal{T}$ is aligned with the skin surface $S_{\mathcal{H}}$ by finding $\rho_{\mathcal{H}}$ that maps each of the template components individually. For this, we rely on a hybrid approach that is largely data-driven but also based on heuristics that ensure anatomic plausibility and avoids intersections.

As the first of the remaining five template meshes, we fit the skin layer by setting 
\begin{equation}\label{eq::skin_warp}
    \hat{S}_{\mathcal{H}} = \text{rbf}_{S_{\mathcal{T}} \rightarrow S_\mathcal{H}}(\hat{S}_{\mathcal{T}}).
\end{equation}
The RBF function is a space warp based on triharmonic radial basis functions \cite{botsch2005real} that is calculated to displace from the template skin surface $S_\mathcal{T}$ to the target $S_\mathcal{H}$ and is then applied to the template skin layer. By construction, the skin layer will be warped semantically consistent and stick close to the targeted skin surface. 

Next, we fit the skull layer $\hat{B}_{\mathcal{H}}$ by invoking a linear regressor $D$ that predicts the distances from the vertices of $\hat{S}_{\mathcal{H}}$ to the corresponding vertices of $\hat{B}_{\mathcal{H}}$ and subsequently minimizing with projective dynamics
\begin{equation}
\begin{aligned}
    \arg\min_{X} & \hspace{1mm}  w_\mathrm{rect} {E}_\mathrm{rect}\of{X, \hat{S}_{\mathcal{T}}} \\
    & + w_\mathrm{dist_2} E_{\mathrm{dist}_2}\of{X, \hat{S}_{\mathcal{H}}, D\of{\hat{S}_{\mathcal{H}}}} \\  
    & + w_\mathrm{curv} E_\mathrm{curv}\of{X, \hat{B}_{\mathcal{T}}}. \\
\end{aligned}
\end{equation}
Here, \begin{equation}
    E_\mathrm{dist_2}\of{X, \hat{S}_{\mathcal{H}}, D\of{\hat{S}_{\mathcal{H}}}} = \sum_{x \in X}\of{\sqrt{\norm{ x - {s_x}}^2} - d_x}^2
    \end{equation} 
ensures that for each vertex $x \in X$ the predicted distances $d_x \in D(\hat{S}_{\mathcal{H}})$ is adhered to.
Apart from  $E_\mathrm{dist_2}$, the same regularizing terms as in Equation \eqref{eq::rect} and \ref{eq::bend} are used. The optimization is initialized with $X = \hat{S}_{\mathcal{H}} - D(S_{\mathcal{H}}) \cdot n(\hat{S}_{\mathcal{H}})$\, where $n(\hat{S}_{\mathcal{H}})$ are area-weighted vertex normals. $D$ is trained on the dataset of \cite{gietzen2019method} (SKULLS) that relates MRI skull measurements to skin surface scans. To ease the learning task, we learn the regressor between PCAs of the skin layers and the skin-to-skull-layer distances. Predicted lengths are set to a minimum value if they fall below a threshold, thus, avoiding skin-skull intersections and numerical issues in downstream physics-based simulations. In Figure \ref{fig:fitting}a, the linear regressor training is visualized.

The muscle layer $\hat{\mathcal{M}}_{\mathcal{H}}$ is fitted by positioning its vertices at the same absolute distances between the corresponding skin and skull layer vertices as in the template, and only passing on a small relative amount $w_\mathrm{rel}$ of the distance changes compared to the template. This approach assumes that the muscle mass in the facial area is only moderately affected by body weight and skull size.

The skull mesh is placed by setting
\begin{equation}
    B_{\mathcal{H}} = \text{rbf}_{\hat{B}_{\mathcal{T}} \rightarrow \hat{B}_{\mathcal{H}}}\of{B_{\mathcal{T}}}.
\end{equation}
The properties of the RBF space warp ensure that the skull mesh remains within the skull layer if the layer is of sufficient resolution. The muscle mesh could be placed in a similar fashion but is not needed in our pipeline any further.

Finally, the soft and muscle tissue tetrahedron meshes $\mathbb{S}_{\mathcal{H}}$ and $\mathbb{M}_{\mathcal{H}}$ can be constructed as described in Section \ref{sec:lhm}. 
On average, the complete fitting pipeline takes only about 500ms for one instance on an Intel i5 12600K processor. Figure \ref{fig:fitting}b) visualizes the overall fitting process.

\textbf{Sampling Expressions} 
In the second part of the pipeline, the actual training instance $\of{{S}_{\mathcal{H}}^\func{exp}, \bar{S}_{\mathcal{H}}^\func{exp}}$  is created. Beginning with deformation transfer \cite{botsch2006deformation} to transfer ARKit\footnote{\url{https://developer.apple.com/}} surface-based blendshapes to $S_{\mathcal{H}}$, we create expression $S_{\mathcal{H}}^\func{exp}$ by linearly blending the blendshapes with weights that are obtained from 8 around 10 minutes long dyadic conversations recorded with a custom iOS app.

\textbf{Training Instance}
Finally, $\bar{S}_{\mathcal{H}}^\func{exp} = \phi\of{\phi^{\dagger}\of{S_{\mathcal{H}}^\func{exp}, \mathbb{S}_{\mathcal{H}}, \mathbb{M}_{\mathcal{H}}, B_{\mathcal{H}}}}$ can be computed. Computing one training instance takes approximately 40 seconds on a AMD Threadripper Pro 3995wx.

\section{Experiments}
Before demonstrating the accuracy and efficiency (Section \ref{sec::nvbs}) of neural volumetric blendshapes, we first evaluate the fitting precision of the LHM (Section \ref{sec::exp::lhmf}) as well as the quality of the proposed physical models (Section \ref{sec::exp::pbs}).

\subsection{LHM Fitting}\label{sec::exp::lhmf}
\begin{figure}[]
\includegraphics[width=1.0\linewidth]{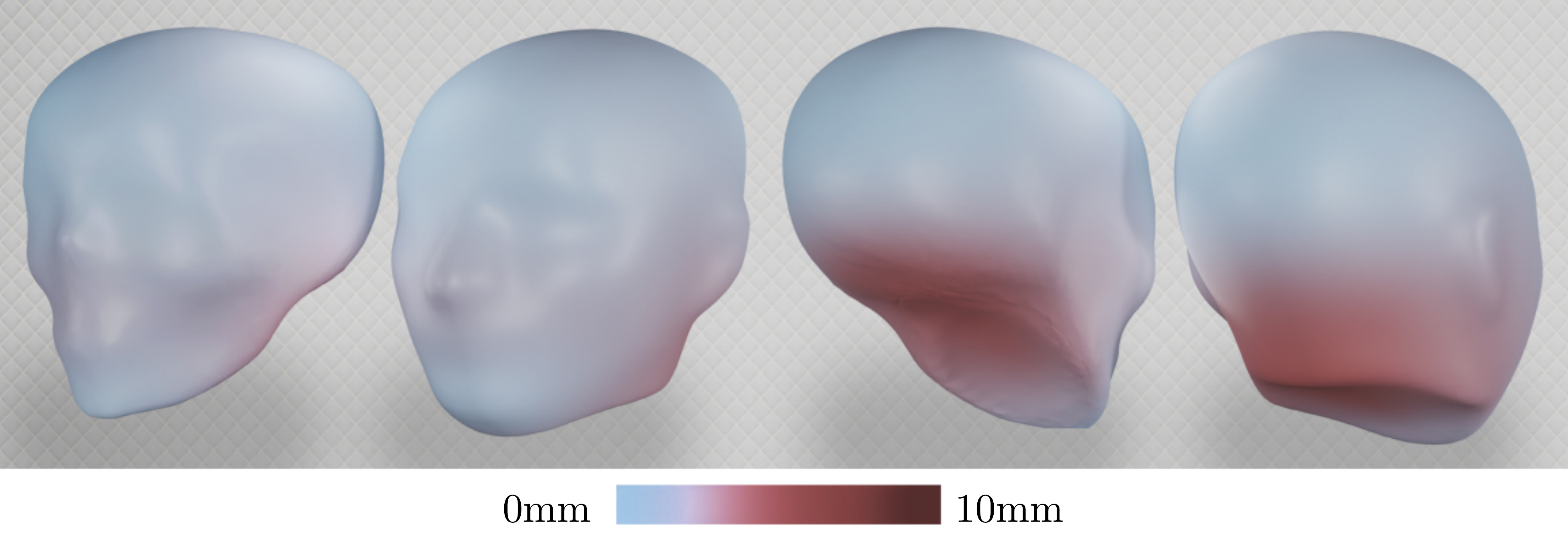}\centering
\caption{The per-vertex mean L2-error of our LHM fitting algorithm in a leave-one-out validation on SKULLS \cite{gietzen2019method}. Larger errors only appear in the back part of the face and have only a minor impact on facial animations.}
\label{img::fitting}
\end{figure}

The fitting of the LHM is mainly composed of the data-driven positioning of the skull layer and the subsequent heuristic fitting of the muscle layer. We evaluate the crucial fitting of the skull layer with the SKULLS \cite{gietzen2019method} dataset. Since this dataset consists of 43 instances only, a leave-one-out validation is performed in which the vertex-wise L2 errors are measured. The results are compared to the multilinear model as originally used for SKULLS in \cite{achenbach2018multilinear}.
\begin{figure*}[t]
\includegraphics[width=1.0\linewidth, height=0.55\linewidth]{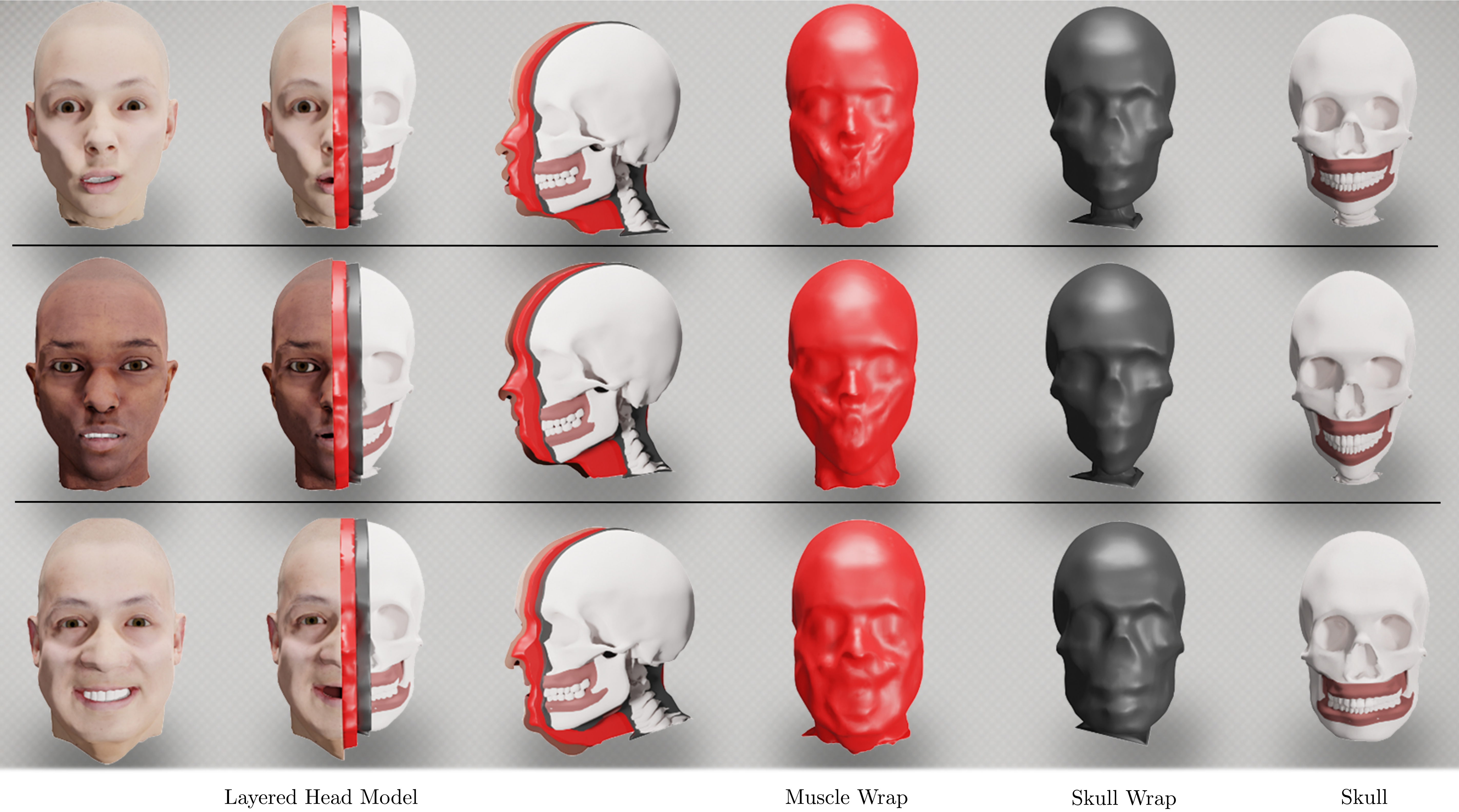}\centering
\caption{Exemplary fits of the LHM components skull wrap, muscle wrap, and skull. The skulls are fitted with a dense position regressor. At this, collisions between the skulls and skins are avoided which increases the numerical stability of downstream physics-based simulations.}
\label{img:preds}
\end{figure*}

Both models cannot achieve a medical-grade positioning with errors between approximately \SI{2}{\milli\meter} and \SI{4}{\milli\meter}. The MLM achieves a higher precision with a mean error of \SI{1.98}{\milli\meter} than our approach that dispositions the skull by \SI{3.83}{\milli\meter} on average. However, the MLM cannot prevent collisions that might crash physics-based simulations. Also, our fitting algorithm produces large errors only in regions that are of less importance for facial simulations as can be seen in Figure~\ref{img::fitting}. The errors are predominately distributed in the back area of the skull, since here the rectangular constraints of our fitting procedure can presumably no longer be aligned well to the skin layer. The following section demonstrates in downstream physics-based simulations that the prediction quality in the frontal face region is sufficiently adequate for detailed facial animations. Figure \ref{img:preds} displays fitting examples.

\subsection{Physics-Based Simulation}\label{sec::exp::pbs}
\begin{figure}[ht]
\includegraphics[width=1.0\linewidth]{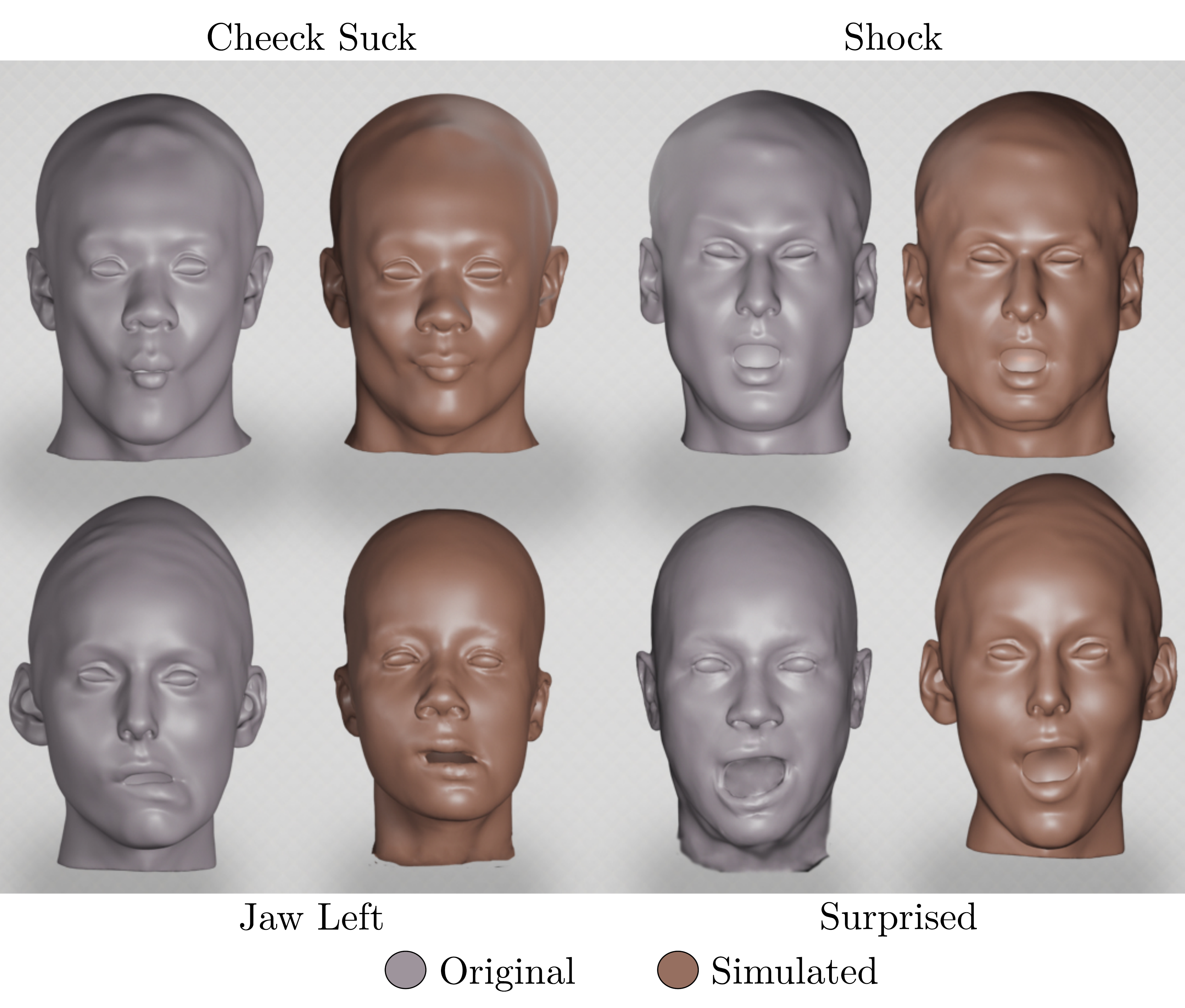}\centering
\caption{Exemplary results of the proposed physics-based facial animation models. In the top row, muscle deformation gradients have been extracted with $\phi^{\dagger}$ and reapplied using $\phi$ to the corresponding neutral head. In the bottom row, the deformation gradients are reapplied to a different identity.}
\label{img::reconst}
\end{figure}
\begin{table}[]
\vspace{4mm}
\centering
\setlength{\tabcolsep}{8pt}
\begin{tabular}{cccccc}

\toprule
$w_\mathrm{{\mathbb{S}_{\mathcal{H}}}}$  & $w_\mathrm{S_{\mathcal{H}}}$  & $w_\mathrm{str}$    & $w_\mathrm{vol}$ & $w_\mathrm{con}$ & $w_\mathrm{inv}$ \\
 $1$& $10$  &  $10^2$ & $10^{3}$ & $10^3$&  $10$  \\ 
$w_\mathrm{{\mathbb{M}_{\mathcal{H}}}}$   & $w_\mathrm{{B_{\mathcal{H}}}}$ & $w_\mathrm{curv}$ & $w_\mathrm{rect}$ & $w_\mathrm{dist}$  & $w_\mathrm{dist_2}$ \\ 
 $1$ & $10$ & $0.1$   &  $1.0$  &  $10.0$  & $10.0$ \\
\bottomrule
\end{tabular}
\caption{The weights used to implement $\phi, \phi^{\dagger}$, and the LHM fitting.}
\label{tab::weights}
\end{table}
\begin{figure*}[]
\centering
\includegraphics[width=0.95\linewidth]{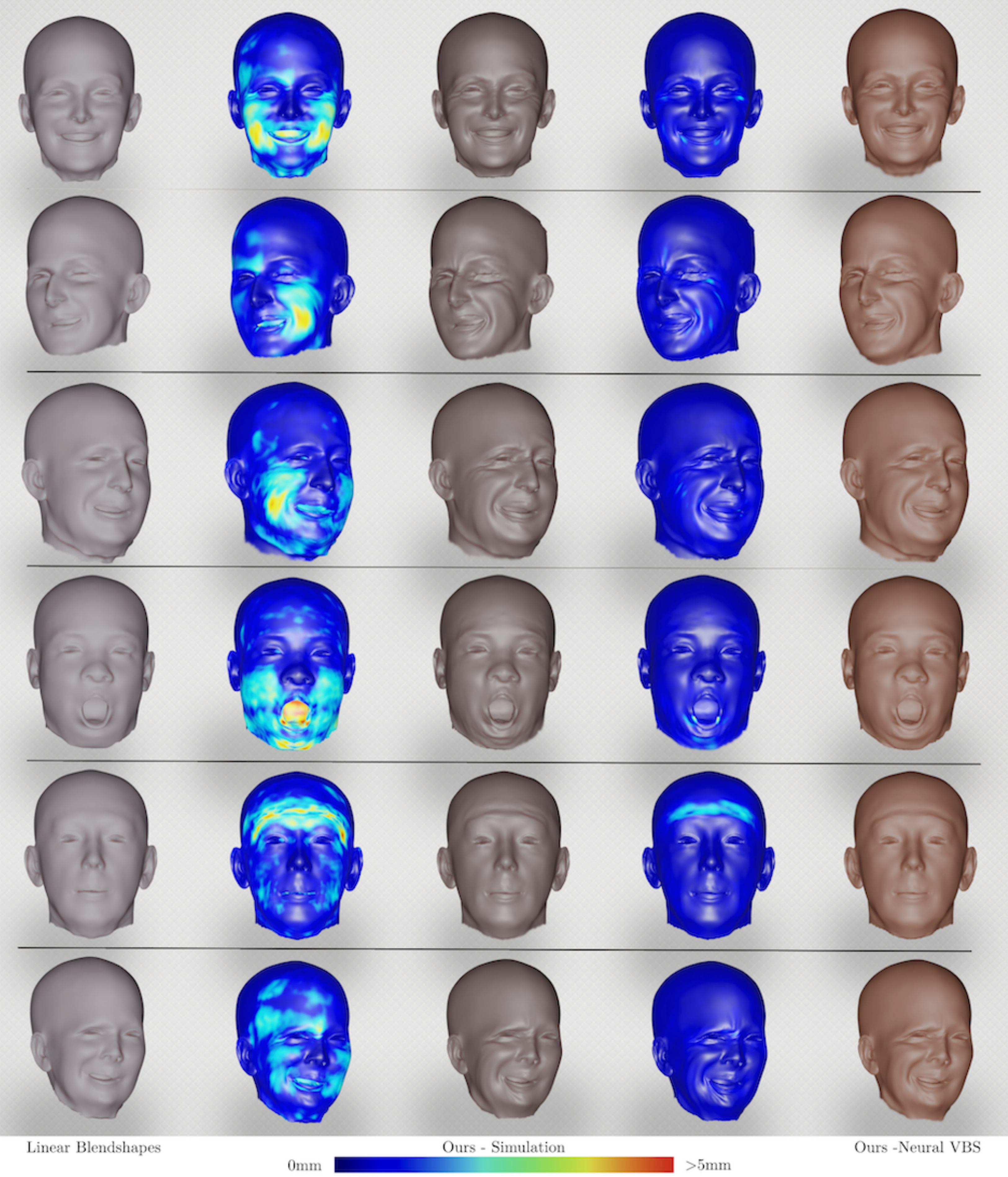}
\vspace{0cm}
\caption{Exemplary test results of our neural volumetric blendshapes for a variety of different head shapes. Besides resolving self-collisions, the physical and anatomical constraints anchored in our approach lead to more realistic and immersive expressions while being computationally efficient.}
\label{img:nvbs_res}
\end{figure*}
To investigate the quality of $\phi^{\dagger}$ and $\phi$ we perform an experiment on nearly 300 high-resolution optical facial expression scans from the proprietary 3DScanstore\footnote{\url{https://www.3dscanstore.com}} dataset. First, we fit the LHM to each of the scans using $\phi^{\dagger}$ and store only the resulting muscle deformation gradients and bone movements. Then, in turn, we apply these to the neutral scans of the corresponding individuals through $\phi$ and simulate the soft tissue as $\phi^{\dagger}$ does. Only if our physic models and the previous LHM fitting operate with reasonable precision, do we expect a low L2-loss between the original expression and the simulated expression as well as visually appealing results.

Across all expression scans, we observe an average L2 reconstruction error of only 1.6 mm. The reproduction quality of the physical models is also reflected in the visual results of Figure \ref{img::reconst}. Further, the accuracy of the LHM fitting investigated in the previous section is again underlined. Additionally, we inspected expression retargeting by applying the extracted muscle and bone transformations to other identities. Again, visual appealing examples can be found Figure \ref{img::reconst}. The weights used to realize $\phi, \phi^{\dagger}$, and the LHM fitting in this and the following experiments are stated in Table $\ref{tab::weights}$.

\subsection{Neural Volumetric Blendshapes}\label{sec::nvbs}
\begin{figure}[t]
\includegraphics[width=1.0\linewidth]{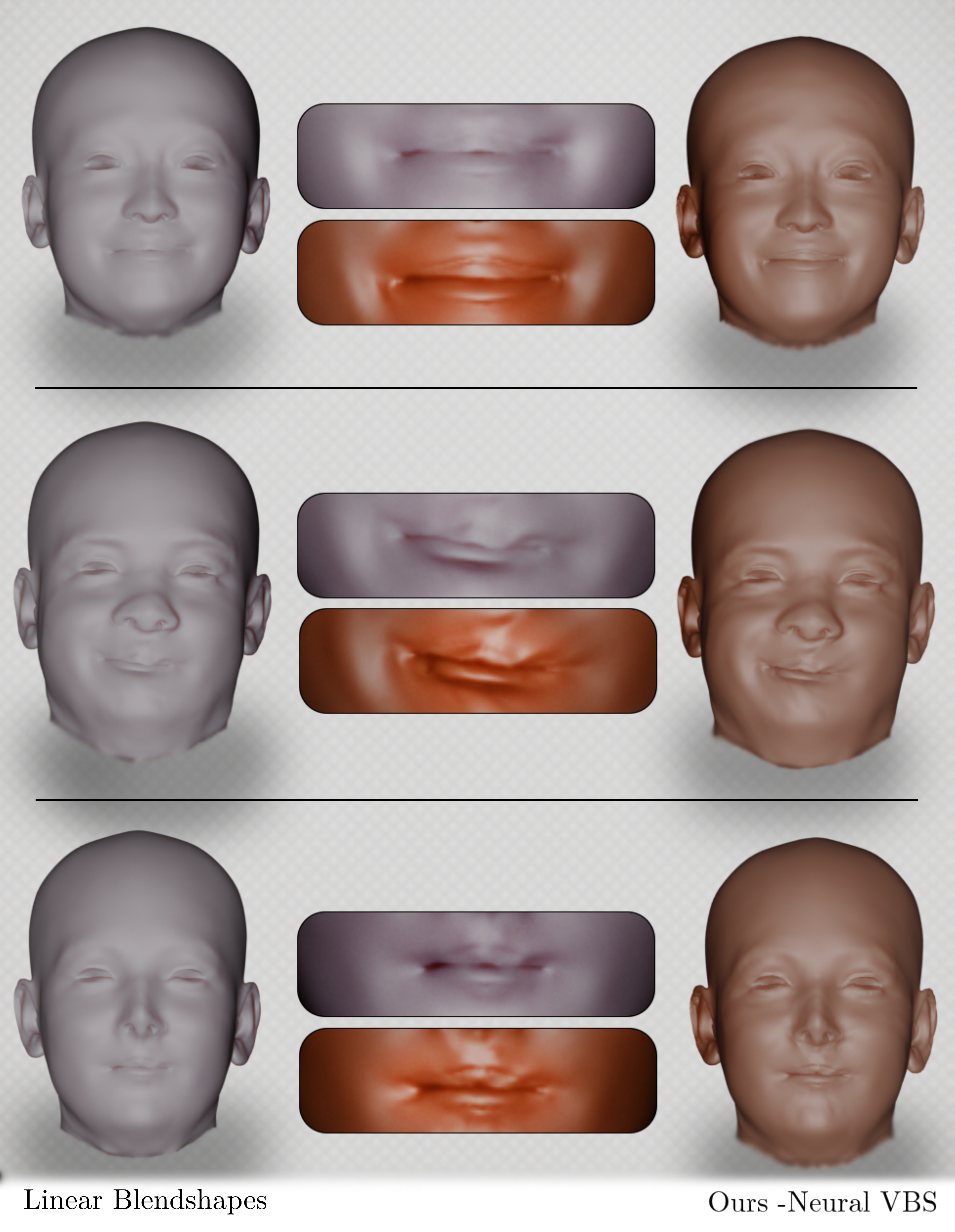}
\vspace{-0.7cm}
\caption{Exemplary results of our collision resolution. Collisions are not only avoided but also simulated in a physically plausible manner.}
\label{img:nvbs_res_cols}
\end{figure}

To train and evaluate $f$, we assemble a dataset of 50000 training and test instances by using the pipeline from Section \ref{sec:pip}. At this, we sample 10000 different head shapes and compute 5 different expressions per head. For training, the Adam optimizer performs 200k update steps with a learning rate of 0.0001. The learning rate is linearly decreased to 0.00005 over the course of training and a batch size of 128 is used. In total, the training specifications result in an approximate runtime of 8 hours on an NVIDIA A6000. The comparatively short training time can straightforwardly be explained by the less noisy training data as usually encountered for image-based deep learning models and the efficient network design. The identities are used 90\% for training and 10\% for testing.

First, we evaluate the efficiency of $f$. The initial runtime of $\phi$ and $\phi^{\dagger}$ in Equation \eqref{eq::eval_nvbs_t} with 6 global projective dynamics updates as in \cite{ichim2016building} is about \SI{950}{\milli\second} on a workstation AMD threadripper pro 3995wx with 128 cores (implemented with ShapeOp \cite{deuss2015shapeop}). Our implementation of $f$ only takes 8ms on a consumer-grade Intel i5 12600K (implemented with PyTorch\footnote{\url{https://pytorch.org}}). Moreover, a forward pass on an NVIDIA RTX3090 is calculated in less than one millisecond. Thus, neural volumetric blendshapes are suitable for realtime applications on weaker hardware setups and also offer advantages when many facial animations are to be executed in parallel on a GPU.

The time advantage alone is not useful if $f$ is not an adequate approximation. We evaluate the approximation quality by measuring the mean L2 error on the test dataset. More precisely, we randomly generate 5 training-test splits of the generated dataset and report the average of the test losses after training on each training set. We achieve an average test error of only \SI{0.13}{\milli\meter}, meaning that we can successfully achieve the approximation target of Equation \eqref{eq::eval_nvbs} and $f$ generalizes well across head shapes and facial expressions. Another open question is the temporal consistency of $f$ which is inherent in physics-based simulations. To this end, we invite the reader to watch the attached video in the supplementary material. Finally, visual results of neural volumetric blendshapes as shown in Figure \ref{img:nvbs_res} exhibit the same improvements towards more realistic facial expressions as much slower physics-based simulations before. These include the resolution of self-intersections, the simulation of wrinkles, the adherence to anatomical boundaries, and the realism-enhancing constraining of volume preservation. Close-up images of resolved self-intersections can be seen in Figure \ref{img:nvbs_res_cols}. The only other approach that to the best of our knowledge can animate detailed faces at this speed, requires only a neutral surface as input, and allows for semantically consistent control, are linear blendshapes paired with a variant of deformation transfer \cite{botsch2006deformation, li2020dynamic, chandran2022local}.  As can be seen in Figure \ref{img:nvbs_res}, too, neural volumetric blendshapes produce far more immersive facial animations.
\section{Conclusion}
In this work, we presented neural volumetric blendshapes, an efficient realization of physics-based facial animations even on con\-sumer-grade hardware. Our approach yields more realistic results than the commonly used linear blendshapes, since artifacts such as volume-loss, self-intersections, or anatomical flaws are avoided and effects like volumetric elasticity and wrinkles are added.
It is also convenient to use, since existing surface-based blendshape rigs can be extended into anatomically and physically plausible animations with almost no integration overhead while keeping the actuation as before.

We aim to improve neural volumetric blendshapes in at least two directions. On the one hand, with an even more accurate anatomical model that represents e.g. trachea and esophagus more precisely. On the other hand, recent results \cite{romero2022contact} show that contact deformations can also be efficiently learned. Since people touch their faces dozens of times \cite{spille2021stop} a day, adding contact-handling for more realistic gestures may improve the immersion significantly.

\newcommand\blfootnote[1]{%
  \begingroup
  \renewcommand\thefootnote{}\footnote{#1}%
  \addtocounter{footnote}{-1}%
  \endgroup
}

\printbibliography

\end{document}